\documentclass[9pt,twocolumn,twoside]{osajnl}

\journal{josab} 

\setboolean{shortarticle}{false}
\usepackage{flushend}

\title{Atmospheric turbulence does not change the degree of polarization of
vector beams}

\author[1,2,$\dagger$]{Zhiwei Tao}
\author[1,2,3,$\dagger$]{Azezigul Abdukirim}
\author[1,2]{Congming Dai}
\author[1,2]{Pengfei Wu}
\author[1,2]{Haiping Mei}
\author[1,2,*]{Yichong Ren}
\author[1,2]{Chuankai Luo}
\author[1,2]{Ruizhong Rao}
\author[1,2]{Heli Wei}

\affil[1]{Key Laboratory of Atmospheric Optics, Anhui Institute of Optics and Fine
Mechanics, Chinese Academy of Sciences, Hefei 230031, China}
\affil[2]{Advanced Laser Technology Laboratory of Anhui Province, Hefei 230037, China}
\affil[3]{Science Island Branch of Graduate School, University of Science and Technology of China, Hefei 230026, China}
\affil[*]{Corresponding author: rych@aiofm.ac.cn}

\begin{abstract}
We propose a novel theoretical framework to demonstrate vector beams whose
degree of polarization does not change on atmospheric propagation. Inspired
by the Fresnel equations, we derive the reflective and refractive field of
vector beams propagating through a phase screen by employing the continuity
of electromagnetic field. We generalize the conventional split-step beam
propagation method by considering the vectorial properties in the vacuum
diffraction and the refractive properties of a single phase screen. Based on
this vectorial propagation model, we extensively calculate the change of
degree of polarization (DOP) of vector beams under different beam parameters
and turbulence parameters both in free-space and satellite-mediated links.
Our result is that whatever in the free-space or satellite-mediated regime,
the change of DOP mainly fluctuates around the order of $10^{-13}$ to $%
10^{-6}$, which is almost negligible.
\end{abstract}

\setboolean{displaycopyright}{true}

\begin{document}

\maketitle

\section{Introduction}

One of the essential properties of light, such as polarization, was proven
to be a paramount resource in many cutting-edge experiments\cite%
{c1,c2,c3,c4,c5,c6,c7}, which offers a proper way to understand the real
physical world and remains to be the theme of much fundamental research
today. The historical survey of understanding the polarization can be dated
back to the underlying theory, bearing Stock's name, in 1852\cite{c8}. It
leads to a straightforward description of degree of polarization (DOP)\cite%
{c9,c10,c11,c12} and provides an elegant geometrical picture to analyze the
impacts of polarization transformations\cite{c13}. However, such theory
cannot describe the random nature of the light emission process. For this
reason, Wolf presents a statistical framework which bridges the connection
between coherence and polarization of random electromagnetic beams\cite%
{c14,c15,c16,c17}.

The question of whether beams whose DOP changes on propagation also arises
on the basis of the recent theory formulated in terms of the cross-spectral
density matrix\cite{c18,c19,c20,c21,c22}. Over past decades, it has been
known that DOP can keep unchanged on propagation, including free-space
channel\cite{c23,c24,c25,c26,c27,c28,c29,c30,c31,c32} or turbulent channel%
\cite{c33,c34,c35,c36,c37}, but it commonly needs to satisfy some specific
conditions. Especially, to investigate the effects of atmospheric turbulence
on DOP, many researchers have focused their efforts primarily on how to
introduce a random phase into the cross-spectral density matrix by
introducing the extended Huygens Fresnel principle\cite{c33,c34,c35,c36,c37}%
. In summary, all of the aforementioned conclusions are obtained through the
coherence theory of light.

\begin{figure}[!b]
\centering
\includegraphics[width=0.8\linewidth]{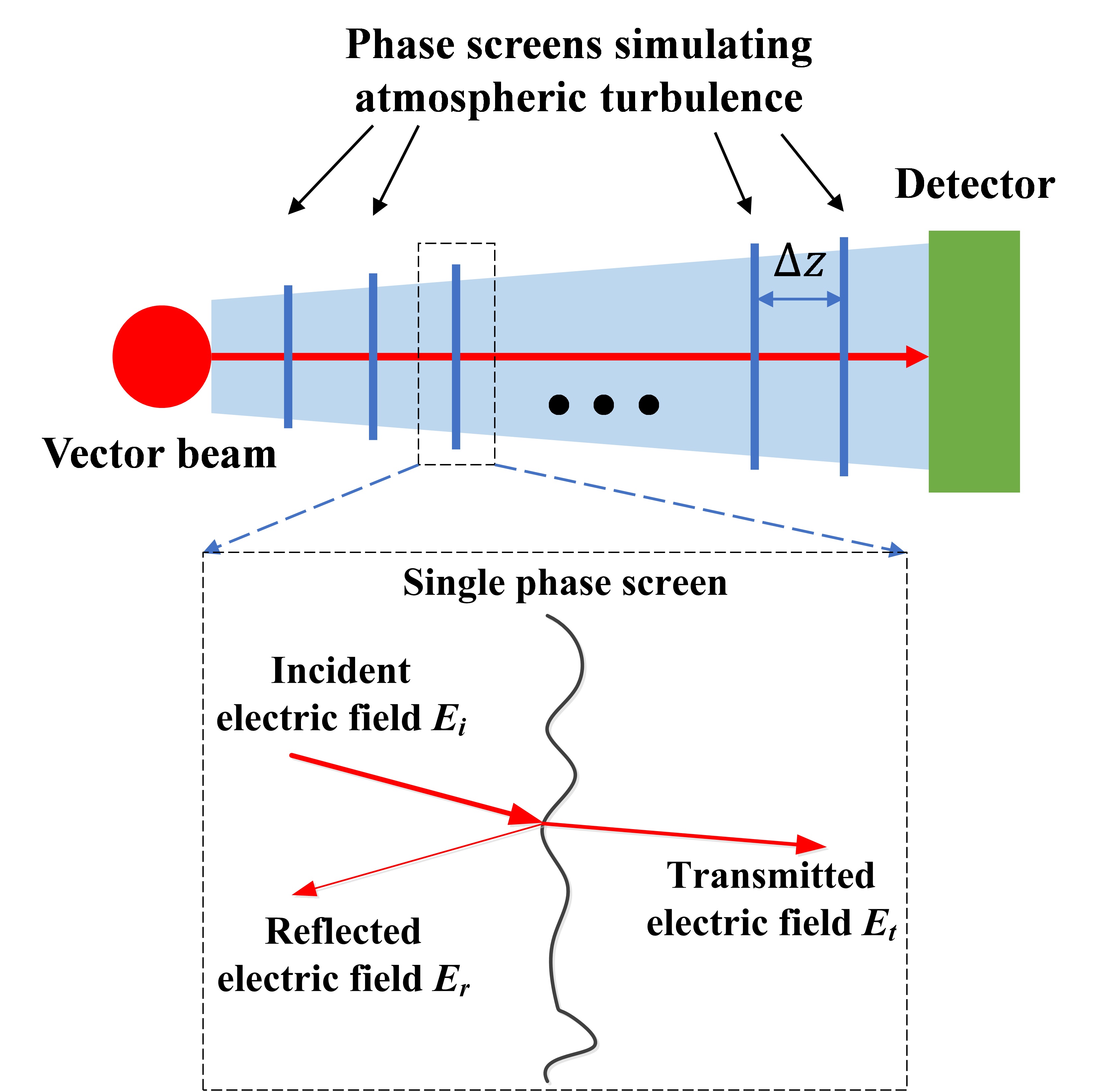}
\caption{The source generates a single vector beam and sends it through a
turbulent atmosphere toward a detector. A turbulent channel is divided into
multiple turbulent cells, each cell introduces a random contribution $%
\protect\varphi $ to the phase, and the effects of reflection and refraction
to the vector beam. The process of vector beams propagating between two
cells is built by the vectorial diffraction theory. The phase screen array
is ladder-shaped distributed because the divergence angle becomes larger as
propagation distance increases.}
\label{fig:1}
\end{figure}

In this article, we propose a novel theoretical framework to demonstrate
vector beams whose DOP does not change on atmospheric propagation, which is
based on the continuity of electromagnetic field\cite{c38,c39} instead of
employing the cross-spectral density matrix. The main difference between our
method and others is that we suggest that light propagation through a random
phase screen is the process of reflection and refraction. Concretly, we
believe that one of the polarized components of light wave falls on to a
boundary between two homogeneous media of different optical properties, it
will be split into two orthogonally polarized components: a component
possess the same polarization with the incident light and another is
orthogonal to the incident one\cite{c40}. After undergoing the modulation of
a series of turbulent cells, it might change the DOP of the incident vector
beam.

\section{Theory}

The conceptual diagram of a vector beam propagating through turbulence that
is depicted by our theoretical model is given in Fig. 1. Since a turbulent
channel can be regarded as a series of equally spaced turbulent cells, the
effects of turbulence on propagation can be splitted into several iterations
of the same modulation. Now, let us begin by recalling the description
scheme of the effects of atmospheric turbulence on scalar beam in general
referring to the split-step beam propagation method\cite{c41}. It means that
each cell introduces a random contribution $\varphi $ to the phase, but
essentially there is no change in the amplitude; besides, intensity
fluctuations build up by diffraction over many cells. It should be noted
that, to agree well with the phase structure function of the Kolmogorov
turbulence model, random phase screen lost low spatial frequencies need to
be compensated using the subharmonic method\cite{c42}.

It is worth highlighting that the conventional theoretical framework of
atmospheric propagation is an incomplete characterization for vector beams,
whereas whether the DOP of beams changes also needs to be investigated
within this method. Moreover, the split-step method will lose its usefulness
when studying the influence of turbulence on a vector beam. Hence, we
further adopt the wave refraction and vectorial diffraction theory\cite%
{c43,c44} together to construct the atmospheric propagation model for a
general polarized vector beam. Without loss of generality, we summarize and
divide the procedure of this model into two steps as follows: Firstly, we
decompose the electric field of a general vector beam into horizontal and
vertical components. Secondly, for each component of the electric field, the
modulation by the random phase and refraction function of the phase screen
and the subsequent diffraction together is repeated several times in the
simulation process. The crucial procedure in the above model that leading to
a change in the DOP of electric field is the
several refractions of multiple phase screen.

To quantify the adverse impact of phase screen on DOP, we report the
refraction of a single random phase screen for the incident electric field $%
\mathbf{E}\left( x,y\right) $ of wave vector $\mathbf{k}$. We assume that $%
E_{x}\left( x,y\right) $ and $E_{y}\left( x,y\right) $ represent the
horizontal and vertical components of $\mathbf{E}\left( x,y\right) $. The
relationship between $\mathbf{k}$ and $\phi \left( x,y\right) $ obeys

\begin{equation}
\mathbf{k}=\nabla \phi \left( x,y\right)  \label{eq1}
\end{equation}%
with $\phi \left( x,y\right) $ denoting the phase of one components of $%
\mathbf{E}\left( x,y\right) $ (i.e., $E_{x}\left( x,y\right) $ or $%
E_{y}\left( x,y\right) $), or more explicitly, $\mathbf{k}=\left( \phi
_{x},\phi _{y},k_{z}\right) $, where $k_{z}^{2}=\left\vert \mathbf{k}%
\right\vert ^{2}-\phi _{x}^{2}-\phi _{y}^{2}$. As shown in Fig. 1, if we
suppose the postive direction of $z$ as the propagation direction, Eq. (\ref%
{eq1}) can be further re-evaluated in terms of incident, transmitted and
reflected components as follows

\begin{eqnarray}
\mathbf{k}^{\left( i\right) } &=&\left( \phi _{x}^{\left( i\right) },\phi
_{y}^{\left( i\right) },\sqrt{\left( 2\pi n_{i}/\lambda \right) ^{2}-\phi
_{x}^{\left( i\right) 2}-\phi _{y}^{\left( i\right) 2}}\right)  \notag \\
\mathbf{k}^{\left( r\right) } &=&\left( \phi _{x}^{\left( r\right) },\phi
_{y}^{\left( r\right) },-\sqrt{\left( 2\pi n_{r}/\lambda \right) ^{2}-\phi
_{x}^{\left( r\right) 2}-\phi _{y}^{\left( r\right) 2}}\right)  \label{eq2}
\\
\mathbf{k}^{\left( t\right) } &=&\left( \phi _{x}^{\left( t\right) },\phi
_{y}^{\left( t\right) },\sqrt{\left( 2\pi n_{t}/\lambda \right) ^{2}-\phi
_{x}^{\left( t\right) 2}-\phi _{y}^{\left( t\right) 2}}\right)  \notag
\end{eqnarray}%
with the subscripts $i$, $r$, $t$ referring to incident, transmitted and
reflected components, respectively, where $\phi _{x}^{\left( l\right) }$ and 
$\phi _{y}^{\left( l\right) }$ ($l=i,r,t$) represent the partial derivative
of one components of $\mathbf{E}^{\left( l\right) }\left( x,y\right) $ ($%
l=i,r,t$) in $x$ and $y$ directions, $n_{l}$ ($l=i,r,t$) denotes the
refractive index of two homogeneous media and $\lambda $ is wavelength. By
means of the orthogonal formula $\mathbf{E}^{\left( l\right) }\left(
x,y\right) \cdot \mathbf{k}^{\left( l\right) }=0$ ($l=i,r,t$), we can
express the incident, transmitted and reflected components of $\mathbf{E}%
\left( x,y\right) $ as

\begin{eqnarray}
\mathbf{E}^{\left( i\right) } &=&\left( E_{x}^{\left( i\right)
},E_{y}^{\left( i\right) },-\left( E_{x}^{\left( i\right) }k_{x}^{\left(
i\right) }+E_{y}^{\left( i\right) }k_{y}^{\left( i\right) }\right)
/k_{z}^{\left( i\right) }\right)  \notag \\
\mathbf{E}^{\left( r\right) } &=&\left( E_{x}^{\left( r\right)
},E_{y}^{\left( r\right) },\left( E_{x}^{\left( r\right) }k_{x}^{\left(
r\right) }+E_{y}^{\left( r\right) }k_{y}^{\left( r\right) }\right)
/k_{z}^{\left( r\right) }\right)  \label{eq3} \\
\mathbf{E}^{\left( t\right) } &=&\left( E_{x}^{\left( t\right)
},E_{y}^{\left( t\right) },\left( E_{x}^{\left( t\right) }k_{x}^{\left(
t\right) }+E_{y}^{\left( t\right) }k_{y}^{\left( t\right) }\right)
/k_{z}^{\left( t\right) }\right)  \notag
\end{eqnarray}%
where $k_{x}^{\left( l\right) }=\phi _{x}^{\left( l\right) }$, $%
k_{y}^{\left( l\right) }=\phi _{y}^{\left( l\right) }$ and $k_{z}^{\left(
l\right) }=\sqrt{\left( 2\pi n_{l}/\lambda \right) ^{2}-\phi _{x}^{\left(
l\right) 2}-\phi _{y}^{\left( l\right) 2}}$ ($l=i,r,t$). Other than that,
the components of magnectic field are obtained by combining the ralation $%
\mathbf{H=}n\frac{\mathbf{k}}{\left\vert \mathbf{k}\right\vert }\times 
\mathbf{E}$ and Eq. (\ref{eq3}), we gives

\begin{eqnarray}
\mathbf{H}^{\left( i\right) } &=&\frac{n_{i}}{\left\vert \mathbf{k}^{\left(
i\right) }\right\vert k_{z}^{\left( i\right) }}\left( 
\begin{array}{c}
-\left[ E_{x}^{\left( i\right) }k_{x}^{\left( i\right) }k_{y}^{\left(
i\right) }+E_{y}^{\left( i\right) }\left( \left\vert \mathbf{k}^{\left(
i\right) }\right\vert ^{2}-k_{x}^{\left( i\right) 2}\right) \right]  \\ 
E_{y}^{\left( i\right) }k_{x}^{\left( i\right) }k_{y}^{\left( i\right)
}+E_{x}^{\left( i\right) }\left( \left\vert \mathbf{k}^{\left( i\right)
}\right\vert ^{2}-k_{y}^{\left( i\right) 2}\right)  \\ 
\left( E_{y}^{\left( i\right) }k_{x}^{\left( i\right) }-E_{x}^{\left(
i\right) }k_{y}^{\left( i\right) }\right) k_{z}^{\left( i\right) }%
\end{array}%
\right)   \notag \\
\mathbf{H}^{\left( r\right) } &=&\frac{n_{r}}{\left\vert \mathbf{k}^{\left(
r\right) }\right\vert k_{z}^{\left( r\right) }}\left( 
\begin{array}{c}
E_{x}^{\left( r\right) }k_{x}^{\left( r\right) }k_{y}^{\left( r\right)
}+E_{y}^{\left( r\right) }\left( \left\vert \mathbf{k}^{\left( r\right)
}\right\vert ^{2}-k_{x}^{\left( r\right) 2}\right)  \\ 
-\left[ E_{y}^{\left( r\right) }k_{x}^{\left( r\right) }k_{y}^{\left(
r\right) }+E_{x}^{\left( r\right) }\left( \left\vert \mathbf{k}^{\left(
r\right) }\right\vert ^{2}-k_{y}^{\left( r\right) 2}\right) \right]  \\ 
\left( E_{y}^{\left( r\right) }k_{x}^{\left( r\right) }-E_{x}^{\left(
r\right) }k_{y}^{\left( r\right) }\right) k_{z}^{\left( r\right) }%
\end{array}%
\right)   \label{eq4} \\
\mathbf{H}^{\left( t\right) } &=&\frac{n_{t}}{\left\vert \mathbf{k}^{\left(
t\right) }\right\vert k_{z}^{\left( t\right) }}\left( 
\begin{array}{c}
-\left[ E_{x}^{\left( t\right) }k_{x}^{\left( t\right) }k_{y}^{\left(
t\right) }+E_{y}^{\left( t\right) }\left( \left\vert \mathbf{k}^{\left(
t\right) }\right\vert ^{2}-k_{x}^{\left( t\right) 2}\right) \right]  \\ 
E_{y}^{\left( t\right) }k_{x}^{\left( t\right) }k_{y}^{\left( t\right)
}+E_{x}^{\left( t\right) }\left( \left\vert \mathbf{k}^{\left( t\right)
}\right\vert ^{2}-k_{y}^{\left( t\right) 2}\right)  \\ 
\left( E_{y}^{\left( t\right) }k_{x}^{\left( t\right) }-E_{x}^{\left(
t\right) }k_{y}^{\left( t\right) }\right) k_{z}^{\left( t\right) }%
\end{array}%
\right)   \notag
\end{eqnarray}%
It is well-known that boundary conditions of electromagnetic field demand
that across the boundary the tangential components of $\mathbf{E}$ and $%
\mathbf{H}$ should be continuous, namely\cite{c38,c39}

\begin{eqnarray}
E_{x}^{\left( i\right) }+E_{x}^{\left( r\right) } &=&E_{x}^{\left( t\right)
},E_{y}^{\left( i\right) }+E_{y}^{\left( r\right) }=E_{y}^{\left( t\right) }
\notag \\
H_{x}^{\left( i\right) }+H_{x}^{\left( r\right) } &=&H_{x}^{\left( t\right)
},H_{y}^{\left( i\right) }+H_{y}^{\left( r\right) }=H_{y}^{\left( t\right) }
\label{eq5}
\end{eqnarray}%
Hence, substituting Eq. (\ref{eq3}) and Eq. (\ref{eq4}) into Eq. (\ref{eq5})
and undergoing a series of algebraic operations, we can immediately
calculate the relationship between $E_{m}^{\left( i\right) }$ and $%
E_{m}^{\left( t\right) }$ ($m=x,y$)%
\begin{eqnarray}
E_{x}^{\left( t\right) } &=&\frac{M_{12}M_{23}-M_{13}M_{22}}{
M_{11}M_{22}-M_{12}M_{21}}  \notag \\
E_{y}^{\left( t\right) } &=&\frac{M_{13}M_{21}-M_{11}M_{23}}{
M_{11}M_{22}-M_{12}M_{21}}  \label{eq6}
\end{eqnarray}%
where

\begin{equation}
M_{11}=M_{22}=\frac{n_{i}}{\left\vert \mathbf{k}^{\left( i\right)
}\right\vert k_{z}^{\left( i\right) }}k_{x}^{\left( i\right) }k_{y}^{\left(
i\right) }+\frac{n_{t}}{\left\vert \mathbf{k}^{\left( t\right) }\right\vert
k_{z}^{\left( t\right) }}k_{x}^{\left( t\right) }k_{y}^{\left( t\right) }
\label{eq7}
\end{equation}

\begin{equation}
M_{12}=F\left( k_{x}^{\left( i\right) },k_{x}^{\left( t\right) }\right)
,M_{21}=F\left( k_{y}^{\left( i\right) },k_{y}^{\left( t\right) }\right)
\label{eq8}
\end{equation}

\begin{equation}
M_{13}=-2\frac{n_{i}}{\left\vert \mathbf{k}^{\left( i\right) }\right\vert
k_{z}^{\left( i\right) }}\left[ E_{x}^{\left( i\right) }k_{x}^{\left(
i\right) }k_{y}^{\left( i\right) }+E_{y}^{\left( i\right) }\left( \left\vert 
\mathbf{k}^{\left( i\right) }\right\vert ^{2}-k_{x}^{\left( i\right)
2}\right) \right]  \label{eq9}
\end{equation}

\begin{equation}
M_{23}=-2\frac{n_{i}}{\left\vert \mathbf{k}^{\left( i\right) }\right\vert
k_{z}^{\left( i\right) }}\left[ E_{y}^{\left( i\right) }k_{x}^{\left(
i\right) }k_{y}^{\left( i\right) }+E_{x}^{\left( i\right) }\left( \left\vert 
\mathbf{k}^{\left( i\right) }\right\vert ^{2}-k_{y}^{\left( i\right)
2}\right) \right]  \label{eq10}
\end{equation}%
with

\begin{equation}
F\left( k_{1},k_{2}\right) =\frac{n_{i}}{\left\vert \mathbf{k}^{\left(
i\right) }\right\vert k_{z}^{\left( i\right) }}\left( \left\vert \mathbf{k}%
^{\left( i\right) }\right\vert ^{2}-k_{1}^{2}\right) +\frac{n_{t}}{
\left\vert \mathbf{k}^{\left( t\right) }\right\vert k_{z}^{\left( t\right) }}%
\left( \left\vert \mathbf{k}^{\left( t\right) }\right\vert
^{2}-k_{2}^{2}\right)  \label{eq11}
\end{equation}

So far, we have quantified the change in the electric field of $\mathbf{E}%
^{\left( i\right) }$ after refraction through the phase screen, which is the
key step to determine whether DOP changes after the electric field is
propagated through turbulence. It should be stressed that when analyzing the
DOP of any component of the vector light, either $E_{x}^{\left( i\right) }$
or $E_{y}^{\left( i\right) }$ of the electric field needs to be set to zero.
Therefore, we can see from Eq. (\ref{eq6}) that even if the incident light
is polarized in a single direction, the propagated light refracted by a
phase screen still has two $x$, $y$ directional polarization components.
More generally, the result of such a modulation is that the $x$ ($y$)
component of the output light is in fact a mixture of the $x$ ($y$)
component produced by the refraction of the $x$ and $y$ direction components
of the incident light through a random phase screen. Finally, we also should
point out that in our theoretical calculations, the fact that DOP does not
change because of diffraction is taken for granted since DOP hardly varies
with the diffraction distance if the various parameters satisfy the
appropriate conditions.

\begin{figure}[!h]
\centering
\includegraphics[width=\linewidth]{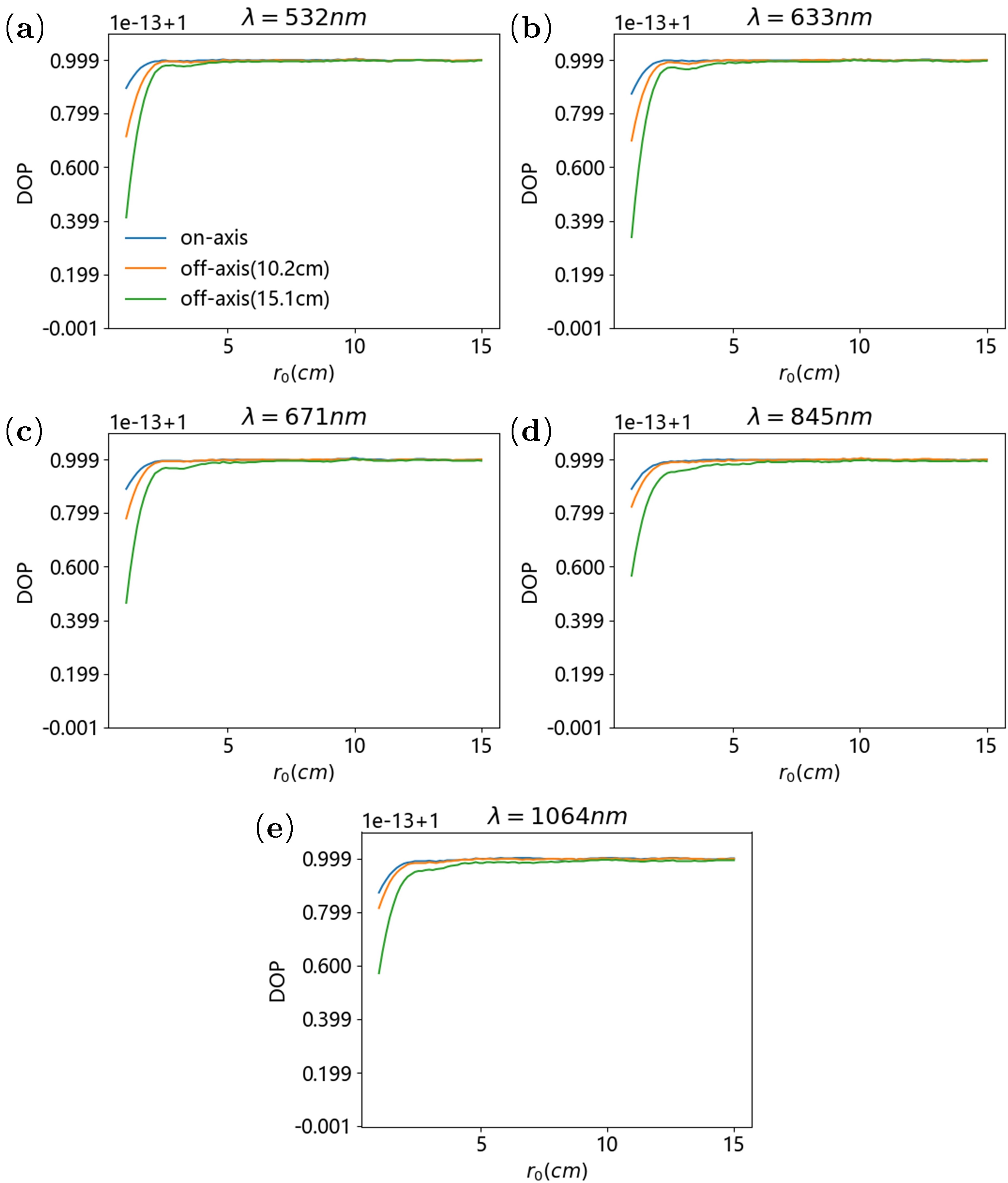}
\caption{DOP of a circularly polarized light as a function of $r_{0}$ for
different values of wavelength: (a) $\protect\lambda=532nm$; (b) $\protect%
\lambda=633nm$; (c) $\protect\lambda=671nm$; (d) $\protect\lambda=845nm$;
(e) $\protect\lambda=1064nm$. All curves in each plot correspond to the DOP
calculated under different off-axis magnitudes.}
\label{fig:2}
\end{figure}

\section{Results}

\subsection{Free-space links}

Fig. 2 illustrates how the DOP of circularly polarized light changes with
respect to the atmospheric coherence length $r_{0}$ under different off-axis
magnitudes, where each plot from upper to bottom represents the results of
different values of wavelength (In the simulation of free-space atmospheric
propagation, we adopt the side length at source plane and receiver plane of $%
30cm$ with a spatial resolution of $1.2mm$ and calculate the DOP averaged
over $1000$ realizations of turbulence). It can be seen that in the strong
turbulence regime, the DOP of circularly polarized light rapidly decreases
with the increase of turbulence strength, whereas in the weak-to-moderate
turbulence regime, DOP almost does not change as $r_{0}$ gradually
decreases. We found that, overall, the change of DOP affected by atmospheric
turbulence mainly fluctuates around the order of $10^{-13}$, which can be
almost ignored. Moreover, we observe that the polarization properties of the
optical field offset from the center of the optical axis are strongly
affected by turbulence. In other words, atmospheric turbulence has smaller
effect on the DOP of on-axis optical field compared to that of off-axis one,
which leads to a reduction of DOP as the off-axis magnitude becomes larger.
Finally, from the comparison between Fig. 2(a) to 2(e), we see that the DOP
at the center of the optical axis hardly varies with the wavelength, yet the
polarization properties of the off-axis optical field may become smaller as
the wavelength increases, which is mainly caused by the fact that a short
wavelength polarized light undergoes a larger effect of atmospheric
turbulence.

\begin{figure*}[!h]
\centering
\includegraphics[width=0.75\linewidth]{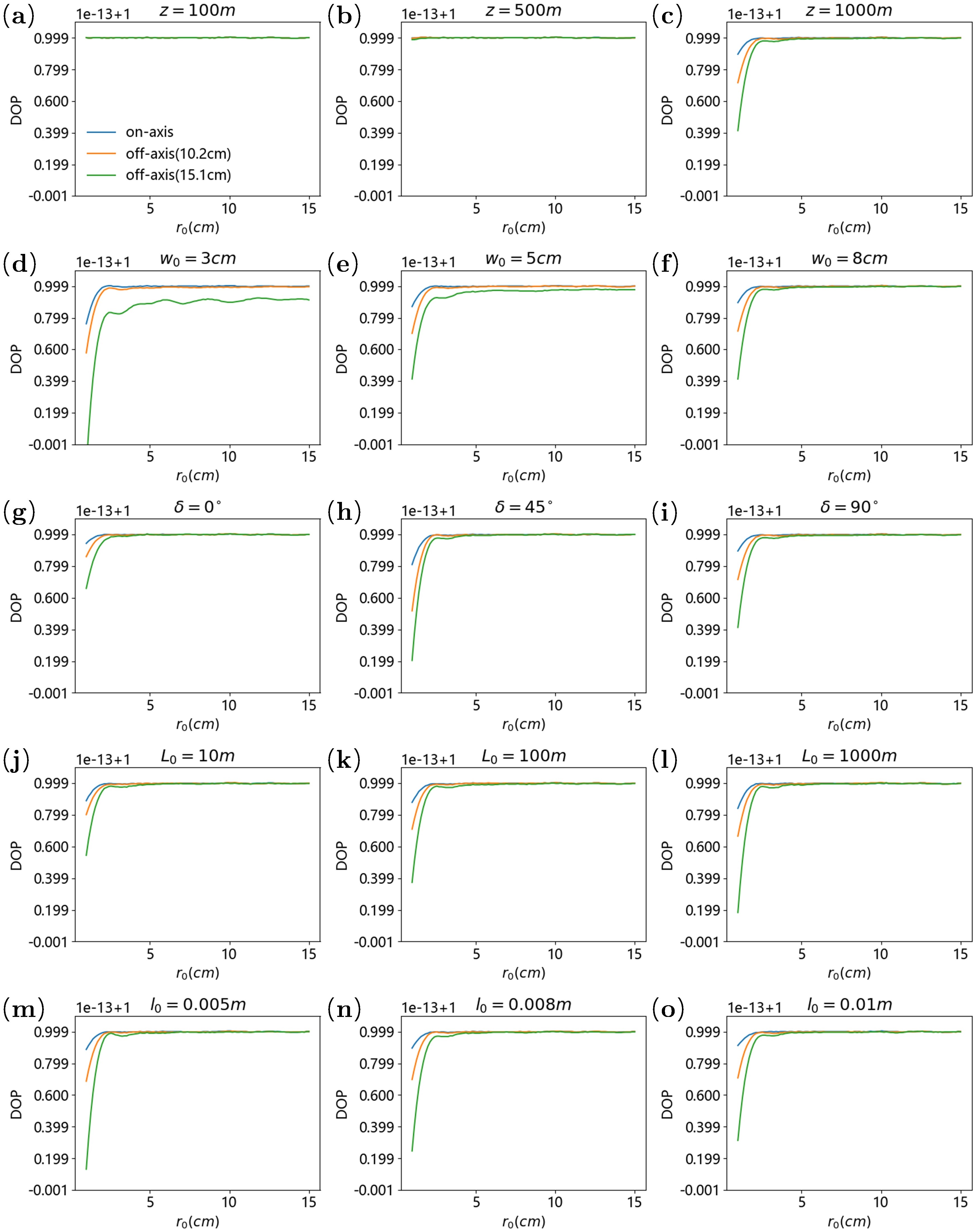}
\caption{DOP of vector beams as a function of $r_{0}$ for different beam
parameters and turbulence parameters, where each row from upper to bottom
represents the results calculated under different propagation distances ((a) 
$z=100m$; (b) $z=500m$; (c) $z=1000m$), beam waists ((d) $w_{0}=3cm$; (e) $%
w_{0}=5cm$; (f) $w_{0}=8cm$), polarization types ((g) $\protect\delta %
=0^{\circ }$; (h) $\protect\delta =45^{\circ }$; (i) $\protect\delta %
=90^{\circ }$), outer scales ((j) $L_{0}=10m$; (k) $L_{0}=100m$; (l) $%
L_{0}=1000m$) and inner scales ((m) $l_{0}=5mm$; (n) $l_{0}=8mm$; (o) $%
l_{0}=10mm$), respectively. All curves in each plot correspond to the DOP
calculated under different off-axis magnitudes.}
\label{fig:3}
\end{figure*}

\begin{figure*}[!h]
\centering
\includegraphics[width=0.5\linewidth]{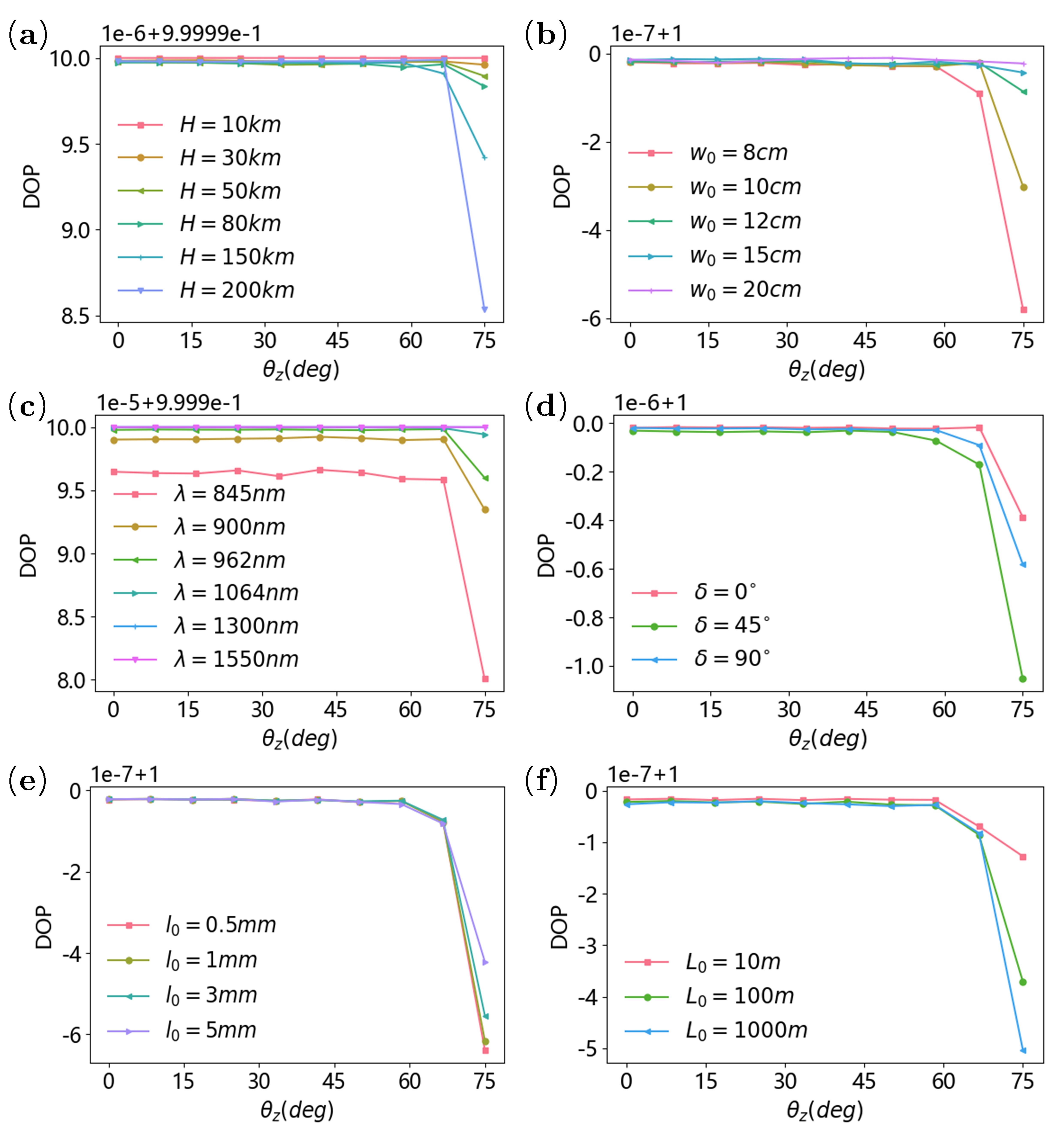}
\caption{DOP of vector beams as a function of $\protect\theta_{Z}$ for
different beam parameters and turbulence parameters: (a) propagation
distance; (b) beam waist; (c) wavelength; (d) polarization types; (e) inner
scale; (f) outer scale. All curves in each plot correspond to the DOP
calculated under different parameter settings.}
\label{fig:3}
\end{figure*}

The DOP of vector beams as a function of $r_{0}$ under different parameter
settings, including beam parameters and turbulence parameters, are shown in
Fig. 3, where each row from upper to bottom represents the results with
different propagation distances $z$ (Fig. 3(a) to (c)), beam waists $w_{0}$
(Fig. 3(d) to (f)), polarization types $\delta $ (Fig. 3(g) to (i)), outer
scales $L_{0}$ (Fig. 3(j) to (l)) and inner scale $l_{0}$ (Fig. 3(m) to
(o)), respectively, where $\delta $ stands for the phase difference between $%
E_{x}^{\left( i\right) }$ and $E_{y}^{\left( i\right) }$. All curves in each
plot correspond to the DOP under different off-axis magnitudes. As depicted
in Fig. 3(a) to 3(c), we observe that DOP starts to decrease with the
increasing propagation distance and turbulence strength, which is likely
because the negative effects of turbulence on vector beams will become
serious as the propagation distance increases. We also notice that in the
short propagation distance or weak turbulence regime, the polarization
properties remain almost constant and is hardly affected by the off-axis
magnitude. Other than that, from the results presented in Fig. 3(d) to (f),
we found that a smaller beam waist of vector beams may be lead to a strong
effect of atmospheric turbulence. The primary reason because vector beams
with a smaller beam waist possesses a larger divergence angle so that it is
more susceptible to atmospheric turbulence\cite{c45}. The other conclusions
about the effects of turbulence strength and off-axis magnitude are the same
as those of Fig. 2. The effects of different polarization types at the
transmitter on DOP with respect to $r_{0}$ are illustrated in Fig. 3(g) to
(i). We notice that in the moderate-to-strong turbulence regime, the DOP of
circularly polarized light may be more vulnerable to atmospheric turbulence
than that of linearly polarized light; besides, when $\delta =45^{\circ }$,
elliptically polarized light has the worst performance when propagating in
atmospheric turbulence. Finally, we reveal the effects of outer scale and
inner scale of turbulence on the DOP of vector beams in Fig. 3(j) to 3(o).
From the comparison between different outer scales and inner scales, we
observe that a large value of $L_{0}$ and a small value of $l_{0}$ may lead
to a significant reduction of DOP, which is partly because inner scale and
outer scale of turbulence forms the lower limit and upper limit of the
inertial range, a smaller value of $l_{0}$ and a larger value of $L_{0}$ are
obtained with the increase of turbulence strength. In the other words, the
decreasing of $l_{0}$ or increasing of $L_{0}$ is equivalent to increase the
number of turbulent cells along the turbulent channel, which causes a vector
beam meets more turbulence during the atmospheric propagation.

\subsection{Satellite-mediated links}

After introducing the DOP of vector beams changes under different parameter
settings in the free-space links, we now turn our attention to the
satellited-mediated links. We further investigate whether the DOP of vector
beams changes in the vertical atmospheric links, which can seen as a poster
child of atmospheric propagation in long-distance and nonuniform turbulent
links. The turbulence strength within a satellite-mediated atmospheric
channel can be described by the refractive index structure parameter $%
C_{n}^{2}$ as a function of altitude $h$ (unlike the calculation of
satellite-mediated atmospheric propagation, we keep $C_{n}^{2}$ constant in
free-space links). We describe $C_{n}^{2}\left( h\right) $ by the widely
used Hufnagel-Valley (HV) model\cite{c46} and adjust the distribution of
turbulent cells by following the so-called rule of equivalent Rytov-index
interval phase screen. More details about this rule and the turbulent link
modeling used to perform the satellite-mediated calculation is further
discussed in the Appendix. Notably, we adjust the side length at source
plane and receiver plane with a specific value according to the propagation
distance and beam waist during the satellite-mediated simulations because of
the divergence properties of the vector beams propagating through
atmospheric turbulence (e.g., when the propagation distance increases from $%
10km$ to $200km$, we set the side length at receiver plane from $1m$ to $3m$%
).

In Fig. 4, we illustrate the variation curves of the on-axis DOP of vector
beams propagating through satellite-mediated links as a function of zenith
angle $\theta _{Z}$ under different parameters (averaged over $1500$
realizations of turbulence). These parameters are the same as in Fig. 3. All
curves in each plot correspond to the on-axis DOP under different parameter
settings. We see in Fig. 4 that, overall, the DOP of vector beams remains
almost unchanged with respect to $\theta _{Z}$ except when $\theta _{Z}$
becomes larger, however, the change of DOP in satellite-mediated links
affected by atmospheric turbulence mainly fluctuates around the order of $%
10^{-5}$ to $10^{-7}$, which is negligible, but is more affected compared to
the results achieved from free-space atmospheric propagation. As depicted in
Fig. 4(a), we compare the on-axis DOP under different propagation distances
(expressed as $H$). We found that in the large $\theta _{Z}$ regime, the DOP
of long-distance propagation shows a significant reduction compare to that
of short-distance one, which indicates that the DOP of on-axis optical field
is gradually affected by atmospheric turbulence as the increase of
propagation distance. In Fig. 4(b), we move our concern to the circumstance
of different beam waists. We clearly observe the same conclusion obtained in
the free-space links where the DOP of vector beams possessing a larger beam
waist outperforms that of vector beams with a smaller one and once again
verify the the primary reason that a smaller beam waist has a larger
divergence angle so that it is more vulnerable in atmospheric turbulence
whatever in free-space or satellite-mediated links. In Fig. 4(c), we
calculate the DOP of a circularly-polorized vector beam propagating through
vertical atmosphere under different values of wavelength. We clearly observe
that the polarization properties is deeply affected by atmospheric
turbulence with the decreasing wavelength (we use the word "deeply" to
describe because the effect of wavelength on DOP varies in several orders of
magnitude). In addition, we notice that the DOP is more affected by the
wavelength for a larger $\theta _{Z}$, which may be caused by the combined
effect of propagation distance and wavelength. We investigate the effects of
different polarization types at the transmitter on DOP with respect to $%
\theta _{Z}$ in Fig. 4(d). We can easily achieve the same conclusion
obtained in Fig. 3(g) to (i), namely, the elliptically polorized vector beam
is more fragile compared to linearly and circularly polarized light. In a
word, it should be emphasized that vector beams with different polarization
types remain almost unaffected by atmospheric turbulence even in the
satellite-mediated links. Fig. 4(e) and (f) plot the DOP of vector beams
propagating in non-Kolmogorov turbulence as a function of $\theta _{Z}$ for
different values of $l_{0}$ and $L_{0}$, where the calculation is shown for
a distance of $150km.$ It can be seen that the smaller $l_{0}$ or the larger 
$L_{0}$ will always lead to a vector beam meet more turbulence either in the
free-space or satellite-mediated propagation regime, for the reason we have
already explained an intuitive way in the previous section.

\section{Conclusions}

In this paper, we have studied propagation of vector optical beams inside
atmospheric turbulence, taking into account the change of polarization
properties both in free-space links and satellite-mediated links. Unlike
previous researches formulated by the cross-spectral density matrix\cite%
{c23,c24,c25,c26,c27,c28,c29,c30,c31,c32,c33,c34,c35,c36,c37}, we propose a
novel propagation model for a generally polarized vector beam in analogy
with the well-known split-step beam propagation method. The main idea behind
our method is that we consider the process of reflection and refraction
exerted by a phase screen, which is derived on the generalization of the
continuity of electromagnetic field. Additionally, we employ the vectorial
diffraction formula to describe the vacuum diffraction between two phase
screens. After making such revisions to the conventional propagation model,
we investigate the change of polarization on atmospheric propagation under
different parameters and parameter settings. It is found that the changes of
DOP in free-space links and satellite-mediated links are mainly surrounded by the
order of $10^{-13}$ and $10^{-6}$ and can nearly be ignored. Our results
further confirmed vector beams whose DOP does not change on atmospheric
propagation and will be useful for free-space optical communications and
quantum communications.

\appendix

\section{Satellite-mediated turbulent link modeling}

We divide the satellite-mediated link into $N$ turbulent cells bounded by
specific altitudes $h_{i}$ with $i$ ranging from $1$ to $N$ (note that
turbulent cells are aranged from lower altitudes to higher altitudes). The
altitude $h_{i}$ of each turbulent cell is calculated by the rule of
equivalent Rytov-index interval phase screen (ERPS). For convenience of
presentation, we summarize and divide the procedure of ERPS's execution into
four steps as follows (we employ the Rytov index $\sigma _{R}^{2}\left(
\Delta h_{i}\right) $ to characterize the scintillation between two
turbulent cells, see more detailed reasons in Ref. \cite{c47}):

1) We set the constant $c$ such that the Rytov index of two adjacent phase
screens is equal to $c$ (i.e., $\sigma _{R}^{2}\left( \Delta h_{i}\right)
\equiv c$, where $\Delta h_{i}=h_{i+1}-h_{i}$).

2) We calculate the altitude of the first phase screen based on $%
C_{n}^{2}\left( 0\right) $ (represent the near-surface refractive index
structure parameter) and the Rytov equation: $\sigma _{R}^{2}\left( \Delta
h_{1}\right) =1.23C_{n}^{2}\left( 0\right) k^{7/6}\left( \Delta h_{1}\right)
^{11/6}\equiv c$, which is employed to set the initial value.

3) We calculate the spacing $\Delta h_{i}$ by using $C_{n}^{2}\left(
h_{i}\right) $ and solving the identity: $1.23C_{n}^{2}\left( h_{i}\right)
k^{7/6}\left( \Delta h_{i}\right) ^{11/6}\equiv c$.

4) We repeat step 3) several times until the sum of the spacing of phase
screen equals the total propagation distance (i.e., we should decide whether 
$\sum_{i=1}^{N}\Delta h_{i}$ is equal to $H$. If not, we repeat step 3; if
yes, we terminate the loop).

So far, we have determined the exact number of $N$ and obtained the specific
altitudes of $h_{i}$. However, it is worth emphasizing that the above
calculation is performed assuming the condition that $\theta _{Z}=0$. If $%
\theta _{Z}\neq 0$, the specific altitudes of turbulent cells should be
adjusted to $h_{1}\sec \theta _{Z}$, $h_{2}\sec \theta _{Z}$, $\cdots $, $%
h_{N}\sec \theta _{Z}$.

Finally, we realize the corresponding random phase screens by employing the
von-Karman spectrum of refractive-index fluctuation\cite{c48} and the
well-known subharmonic-conpensation-based fast-Fourier-transform algorithm%
\cite{c41,c42,c49,c50}, which is implemented on the python library named
AOtools\cite{c51}.\\

\noindent\textbf{Funding.} HFIPS Director's Foundation (YZJJ2023QN05); National Key R$\&$D Program of China (2019YFA0706004); National Natural Science Foundation of China (11904369); National Key R$\&$D Program of Young Scientists (SQ2022YFF1300182); Youth Innovation Promotion Association of Chinese Academy of Sciences (2022450).\\

\noindent\textbf{Acknowledgment.} We would like to thank the anonymous
reviewers for their valuable comments, which significantly improves the
presentation of this paper.\\

\noindent\textbf{Disclosures.} The authors declare no conflicts of interest.\\

\noindent$^\dagger$These authors contributed equally to this article.

\end{document}